\RequirePackage{lineno}
\documentclass[preprint,aps,prc,showpacs,superscriptaddress,preprintnumbers,floatfix,nofootinbib]{revtex4}
\usepackage{epsfig,graphics}
\usepackage{graphicx}
\usepackage{dcolumn}
\usepackage{bm}
\usepackage{amsmath}
\usepackage[usenames]{color}
\usepackage{ulem} 

\begin{document}

\title{Investigating the quark flavor dependence of the chiral magnetic effect with a multiphase transport model}

\author{Ling Huang}
\affiliation{Institute of Particle and Nuclear Physics, Henan Normal University, Xinxiang 453007, China}
\affiliation{Shanghai Institute of Applied Physics, Chinese Academy of Sciences, Shanghai 201800, China}
\author{Chun-Wang Ma}
\email[]{machunwang@126.com}
\affiliation{Institute of Particle and Nuclear Physics, Henan Normal University, Xinxiang 453007, China}
\author{Guo-Liang Ma}
\email[]{glma@sinap.ac.cn}
\affiliation{Shanghai Institute of Applied Physics, Chinese Academy of Sciences, Shanghai 201800, China}


\begin{abstract}
Because the properties of the QCD phase transition and the chiral magnetic effect (CME) depend on the number of quark flavors ($N_{f}$) and quark mass, relativistic heavy-ion collisions provide a natural environment to investigate the flavor features if quark deconfinement occurs. We introduce an initial two-flavor or three-flavor dipole charge separation into a multiphase transport (AMPT) model to investigate the flavor dependence of the CME. By taking advantage of the recent ALICE data of charge azimuthal correlations with identified hadrons, we attempt to disentangle two-flavor and three-flavor CME scenarios in Pb+Pb collisions at 2.76 TeV. We find that the experimental data show a certain potential to distinguish the two scenarios, therefore we further suggest to collect more data to clarify the possible flavor dependence in future experiments. 

\end{abstract}


\maketitle

\section{Introduction}
\label{sec:intro}

Relativistic heavy-ion collisions serve as a very good laboratory to investigate the intrinsic characters of fundamental QCD interactions, because they can convert confined hadronic matter with chiral symmetry breaking into deconfined partonic matter with chiral symmetry restoration by varying temperature and baryon chemical potential~\cite{Karsch:2001cy,Ding:2015ona,Bazavov:2014pvz}. In the past decade, plenty of experimental results announce that a strongly-coupling quark gluon plasma (sQGP) has been created by relativistic heavy-ion collisions at the BNL Relativistic Heavy-Ion Collider (RHIC) and the CERN Large Hadron Collider (LHC)~\cite{Adams:2005dq,Adcox:2004mh,Aamodt:2008zz}. As the same time, an extremely large magnetic field can be produced in the early state of relativistic heavy-ion collisions due to the fast relative motion of two positive-charged nuclei~\cite{Bzdak:2011yy,Deng:2012pc}. If there occurs any topology-changing transition which results in a chirality asymmetry between left- and right-handed quarks in the early state of heavy-ion collisions, an electric current will be generated along the magnetic field direction, which is the so-called chiral magnetic effect (CME)~\cite{Fukushima:2008xe,Kharzeev:2007jp,Hattori:2016emy}. The experimental measurements of charge azimuthal correlation [$\gamma=\left\langle cos(\phi_{\alpha}+\phi_{\beta} - 2\Psi_{RP}) \right\rangle$=$\left\langle cos(\phi_{\alpha}+\phi_{\beta} - 2\phi_{c}) \right\rangle$/$v_{2,c}$] show some consistent results with the CME expectation~\cite{Abelev:2009ac,Abelev:2009ad,Abelev:2012pa}, though the truth is still under debate because many important background sources can largely contribute to the measured signal~\cite{Schlichting:2010qia,Pratt:2010zn,Bzdak:2009fc,Bzdak:2010fd,Liao:2010nv,Wang:2009kd,Bzdak:2012ia}. But the scientific importance of the CME is undoubted, since the CME is deeply linked to the two intrinsic properties of strong interactions, i.e. chiral symmetry and confinement. In other words, if the CME exists in relativistic heavy-ion collisions, it will be a strong evidence of deconfinement and chiral symmetry restoration in the hot and dense strong-interacting QCD matter~\cite{Kharzeev:2007jp}. 

Lattice QCD calculations reveal that the properties of the QCD phase transition depend on the number of quark flavors ($N_{f}$) and their masses~\cite{Karsch:2001cy,Ding:2015ona}. The critical temperature ($T_{c}$) of deconfinement (or chiral symmetry restoration) is different between two-flavor ($N_{f}$=2) QCD and  three-flavor  ($N_{f}$=3) QCD. It is even argued that there could be a flavor hierarchy for the deconfinement transition of QCD matter~\cite{Bellwied:2013cta}, though the $T_{c}$ difference could be very limited~\cite{Ding:2015ona}. The QCD phase diagram of three-flavor QCD sheds light on the fact that the order of phase transition strongly depends on light quark ($u$,$d$) masses and a strange ($s$) quark mass~\cite{Karsch:2001cy,Ding:2015ona}. At the same time, the equation of state of QCD matter must have dependences on the number of quark flavors ($N_{f}$) and their masses~\cite{Karsch:2001cy,Bazavov:2014pvz}. On the other hand, the CME can carry the information about the number of quark flavors ($N_{f}$), because the CME current is proportional to a trace coefficient related to anomalous triangle diagram which gives a coefficient of 5/9 to the two-flavor CME but 2/3 to the three-flavor CME~\cite{Kharzeev:2010gr}. It results in a relative CME current difference of 1/6 between two-flavor and  three-flavor cases. However, because the traditional experimental observable $\gamma$ of charge azimuthal correlation includes all inclusive charged hadrons, it limits ones to measure the relative difference. Recently, the ALICE collaboration measured the identified hadron triggered charge azimuthal correlation, e.g. kaon-hadron correlation~\cite{Onderwaater:2015toa}, which enable us to access the flavor properties of the CME. Another advantage of kaon-hadron correlation is that it can avoid the contamination from $\rho$ meson decays, which presumedly have a significant influence on the CME observable~\cite{Zhao:2017nfq}.  In this work, we will implement a multiphase transport (AMPT) model to investigate the flavor dependence of the CME. Recalling our previous work, we imported the initial CME-like dipole charge separation to the AMPT model with a constant separation percentage, and found that though the original AMPT model can reproduce 60-70\% magnitude of  the CME observable $\gamma$ in Au+Au collisions at 200 GeV, an initial charge separation percentage of $\sim$ 10\% is needed to closely match the STAR data, and demonstrated that  final state interactions strongly suppress the initial CME effect~\cite{Ma:2011uma}. Based on those results, we improve our method to study the $\gamma$ correlators in Pb+Pb collisions at 2.76 TeV in this work, by taking the two following improvements into account. First, we use a centrality-dependent charge separation percentage which is assumed to be proportional to the magnitude of magnetic field in Pb+Pb collisions, which is more reasonable for simulating the CME. Second, we consider both two-flavor and three-flavor cases for the initial charge separation to mimic the two-flavor and three-flavor CME effect, while we only considered two flavors of quarks ($u$ and $d$) in our previous work~\cite{Ma:2011uma}.

This paper is organized as follows. We give an introduction of our model and how to introduce the two-flavor and three-flavor dipole charge separations in Sec.~\ref{sec:model}. In Sec.
~\ref{sec:results}, we present our simulation results and expand some discussions. Finally, we summarize in Sec.~\ref{sec:summary}.

\section{The AMPT Model}
\label{sec:model}
We implement the AMPT model with string melting mechanism in this study~\cite{Lin:2004en}. The AMPT model is a Monte Carlo hybrid transport model, including four main stages of relativistic heavy-ion collisions, i.e., initial condition, parton cascade, hadronization, and hadronic rescatterings. The initial condition, including the spatial and momentum distributions of minijet partons and soft string excitations, is obtained from HIJING model~\cite{Wang:1991hta,Gyulassy:1994ew}, where minijets and soft string excitations are fragmented into hadrons according to the Lund string fragmentation~\cite{Sjostrand:2000wi}. The string melting mechanism can convert all hadrons to quarks according to the flavor and spin structures of their valence quarks, which mimics the QCD deconfinement that results in a formation of a quark and anti-quark plasma.  Next the plasma starts the parton evolution which is simulated by Zhang's parton cascade (ZPC) model~\cite{Zhang:1997ej}, where the perturbative QCD partonic cross section is controlled by the strong coupling constant and Debye screening mass. For hadronization, the  model recombines the freezeout partons to hadrons through a naive coalescence model~\cite{He:2017tla,Chen:2008vr}. The evolutions of the subsequent hadronic phase is then described by a relativistic transport (ART) model~\cite{Li:1995pra}. Recently, Ma and Lin found that the string melting AMPT model with an universal setting of tuned parameters can reasonably reproduce dN/dy, $p_{\rm T}$-spectra, azimuthal anisotropies $v_n$, and longitudinal decorrelation in A+A collisions at both RHIC and LHC energies~\cite{Ma:2016fve}. Therefore, we adopt the same setting of parameters in this work. However, because some hadronic reaction channels from the ART part do not conserve electric charge, we turn off the hadron evolution but only keep resonance decays to ensure charge conservation in an event-by-event manner to study the charge-dependent CME observables. 

\begin{figure}
\includegraphics[scale=0.6]{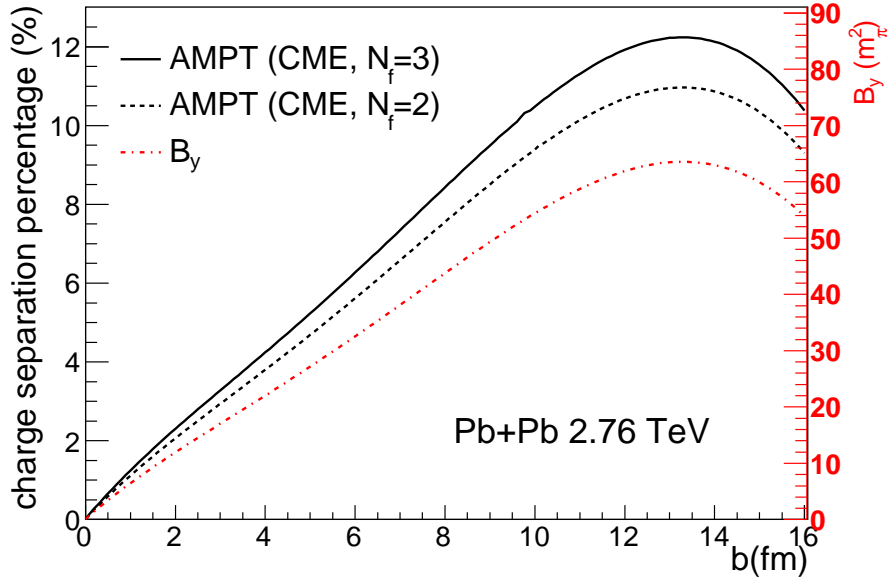}
\caption{(Color online) The impact parameter $b$ dependences of magnetic field $B_y$ (see right $y$ axis, dot-dash) and initial charge separation percentages (see left $y$ axis) for two-flavor CME (dash) and three-flavor CME (solid) in Pb+Pb collisions at 2.76 TeV. The magnetic field $B_{y}$ is taken from Ref.~\cite{Deng:2012pc}.}
\label{fig-1}
\end{figure}

To mimic the CME current, we need to introduce an initial charge separation into the initial condition of the AMPT model, since the motions of charges are not directionally separated but isotropic in the original version of AMPT model. In order to separate a percentage of the initial charges, we use the same method invented by Ma and Zhang in Ref.~\cite{Ma:2011uma}. For a given flavor of quarks, let us say $u$ quark, we switch the $p_y$ values of a percentage ($f$) of the downward moving $u$ quarks with those of the upward moving $\bar{u}$ quarks in such a way that the total momentum is conserved, and likewise for $\bar{d}$ and $d$ quarks and $\bar{s}$ and $s$ quarks. The $y$ axis is perpendicular to the reaction plane ($x$ axis) with the $z$ axis being the incoming direction of one nucleus. The percentage $f$ can be defined as
\begin{equation}
f = \frac{N_{\uparrow(\downarrow)}^{+(-)}-N_{\downarrow(\uparrow)}^{+(-)}}{N_{\uparrow(\downarrow)}^{+(-)}+N_{\downarrow(\uparrow)}^{+(-)}} \propto B_y,
 \label{eq-f}
\end{equation}
where $N$ is the number of quarks, + and - denote positive-charged and negative-charged, and $\uparrow$ and $\downarrow$ represents upward and downward for momentum $p_y$, where we assume that magnetic field is in the $y$ direction. We can introduce two-flavor and three-flavor CME-like electric currents including and excluding the separation of strange quarks ($s$ and $\bar{s}$). For all cases, we assume that all kinds of flavors of quarks share a same value of charge separation percentage for simplicity. In principle, the charge separation percentage due to the CME should be presumably proportional to the magnetic field $B_y$. Since the magnitude of magnetic field is impact parameter($b$)-dependent, the percentage $f$ should depend on $b$ (or centrality bin). To obtain the $b$-dependent percentage, we take the centrality dependence of magnetic field $B_y$ in Pb+Pb collisions at 2.76 TeV from Ref.~\cite{Deng:2012pc}, which is presented by dot-dash curve in Figure~\ref{fig-1}. By assuming that the initial charge percentage obeys Eq.~(\ref{eq-f}),  we use the AMPT model with the initial charge separation to fit the charge azimuthal correlation $\gamma$ from the ALICE data~\cite{Abelev:2012pa} to get the corresponding proportional coefficient for two CME scenarios. Thus, we obtain $b$-dependent initial charge separation percentages, as shown by dash ($N_f$=2) and solid ($N_f$=3) curves in Figure~\ref{fig-1}, which serve as our input for simulating the CME-related observables with the AMPT model. Note that our switchings do not change the initial time of partons, which means that the initial dipole charge separation is introduced around the mean formation time of partons, i.e. 2--3 fm/$c$ in Pb+Pb collisions at 2.76 TeV from the AMPT model, when magnetic field is expected to vanish~\cite{McLerran:2013hla}.

\section{Results and Discussions}
\label{sec:results}
For Pb+Pb collisions at 2.76 TeV, we calculate two kinds of two-particle charge azimuthal correlations, one case is the traditional observable, which is usually called the charge azimuthal correlation $\gamma$, where two particles belongs to unidentified charged hadrons. The other case is the identified hadron triggered charge azimuthal correlation ${c}^{ij}_{\alpha\beta}$, where one of two particles is identified hadron, e.g. kaon, and the other one is an unidentified charged hadron. We will show the two cases in the two following subsections, respectively.

\subsection{Charge azimuthal correlation $\gamma$}
\label{sec:part A}

We calculate the traditional CME observable $\gamma$ through a method similar to that used by the experimentalists~\cite{Abelev:2009ac,Abelev:2009ad,Abelev:2012pa,Voloshin:2004vk}. The charge azimuthal correlation (correlator) $\gamma$ is defined as,
\begin{equation} \label{psi}
\gamma=\left\langle cos(\phi_{\alpha}+\phi_{\beta}-2\Psi_{RP}) \right\rangle,
\end{equation}
where $\phi_{\alpha}$ and $\phi_{\beta}$ represent two charged particles' azimuthal angles, $\alpha$ and $\beta$ denotes their charge which can be "+" or "-", and $\Psi_{RP}$ is the reaction plane angle.

The reaction plane angle $\Psi_{RP}$ is represented by the second order of participant plane $\Psi_{2}$ in our calculations, which can be calculated based on the initial geometry information of partonic system from the AMPT model,
\begin{equation} 
\label{psi}
\Psi_{n}=\frac{1}{n}\left [arctan\frac{\left\langle r^{n}sin(n\varphi)\right\rangle} {\left\langle r^{n}cos(n\varphi)\right\rangle}+\pi \right],
\end{equation}
where $r$ is the displacement of the participating partons from the center of mass, $\varphi$ is the azimuthal angle of the participating partons in the spacial transverse plane, and we take n=2 for $\Psi_{2}$~\cite{Alver:2010gr,Ma:2010dv}.

\begin{figure}
\includegraphics[scale=0.6]{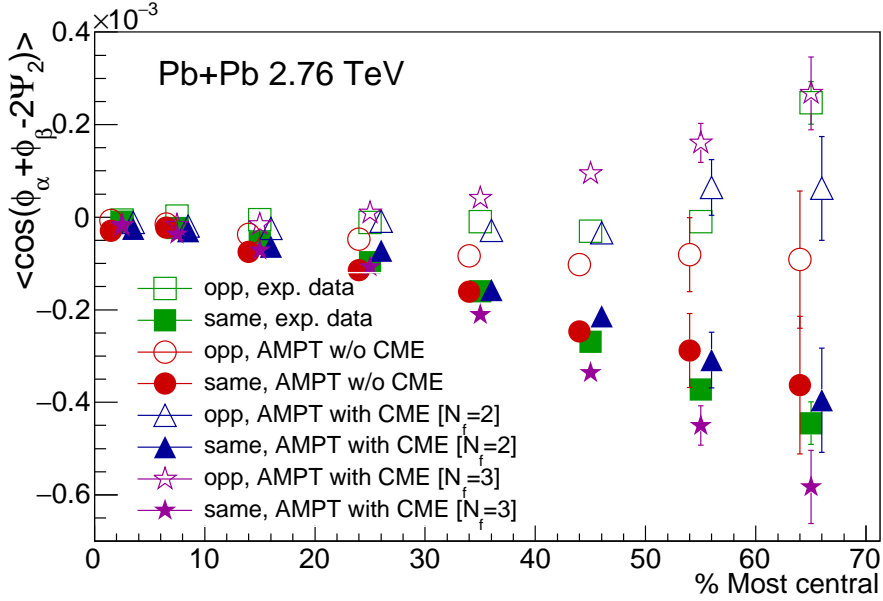}
\caption{(Color online) The centrality dependence of charge-particle azimuthal correlation $\gamma$ in Pb+Pb collisions at 2.76 TeV, where the experimental data are taken from Ref.~\cite{Abelev:2012pa}.
}
\label{fig-2}
\end{figure}

By applying the initial charge separation into the AMPT model, we get the results about $\gamma$ correlator for different cases. In Figure~\ref{fig-2}, square symbol represents experimental data from ALICE~\cite{Abelev:2012pa}, while circle symbol represents the original AMPT results without initial charge separation. We can see that the original AMPT model can basically describe the experimental data for the centrality bin of 0-30\%, but its magnitude is smaller than the experimental data for the centrality bin of 30-70\%, especially for opposite-charge data. This indicates the original AMPT model can serve as a good background model to study the CME, since it can give a main part of $\gamma$ observable even without introducing any initial charge separation. This is mainly because of the advantage of the AMPT model that takes several important background sources into account, including transverse momentum conservation~\cite{Pratt:2010gy,Bzdak:2010fd}, local charge conservation~\cite{Pratt:2010gy}, etc. The triangle symbol represents the results from the AMPT model with initial two-flavor charge separation for $u$ and $d$ quarks. For the centrality bin of 0-60\%, the AMPT results are similar to the experimental data. But for the centrality bin of 60-70\%, it can reproduce same-charge correlation but underestimate opposite-charge one. On the other hand, the star symbol represents the AMPT results with initial three-flavor charge separation for $u$, $d$ and $s$ quarks. We can see that the results from the three-flavor case are close to the experimental data for the centrality bin of 0-30\%. Unfortunately, they overestimate experimental data for the both opposite- and same-charge correlations for the centrality bin of 30-70\%. In general, all three cases can describe the experimental data qualitatively, but none of them can perfectly describe both opposite- and same-charge experimental data for all centrality bins.

\subsection{Identified hadron triggered charge azimuthal correlation $\it{c}^{ij}_{\alpha\beta}$}
\label{sec:part B}
To see the difference between two-flavor CME and three-flavor CME, it is key to see if the strange quarks participate in the CME process. Therefore, it will be very helpful to see the charge azimuthal correlation with a identified hadron as a trigger particle, especially when a charged kaon is the triggered hadron since it consists of a strange quark.  We defined the identified hadron triggered charge azimuthal correlation in the same way as the ALICE collaboration~\cite{Onderwaater:2015toa}:
\begin{equation} \label{psi}
\it{c}^{ij}_{\alpha\beta}(p_T)=\left\langle cos(\phi^i_{\alpha}(p_T)+\phi^j_{\beta}-2\Psi_{RP}) \right\rangle, 
\end{equation}
where $\phi$ represents the azimuthal angle, $i$ denotes an identified hadron and $j$ denotes an unidentified or inclusive charged hadron, $\alpha$ and $\beta$ denotes their charges which can be "+" or "-", and $\Psi_{RP}$ is the reaction plane angle for which we still use $\Psi_{2}$ to represent it. The observable is studied as a function of $p_T$, which is the transverse momentum of the identified hadron.

\begin{figure}
\includegraphics[scale=0.6]{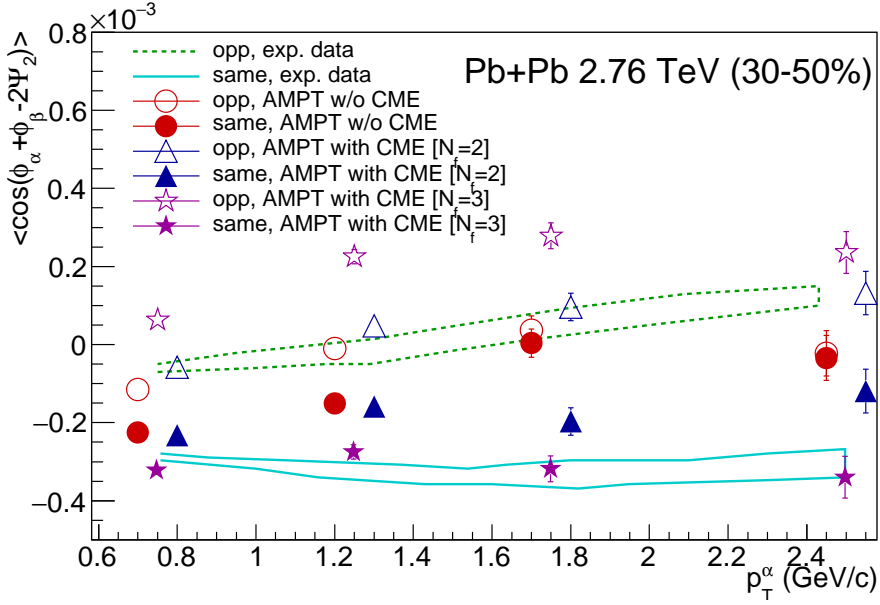}
\caption{(Color online)  The $p_T$ dependence of hadron-hadron charge azimuthal correlation in Pb+Pb collisions (30-50\%) at 2.76 TeV, where two bands represent the ALICE data~\cite{Onderwaater:2015toa}.
}
\label{fig-3}
\end{figure}

To see the difference between two-flavor CME and three-flavor CME, we calculate both hadron-hadron and kaon-hadron correlations. Figure~\ref{fig-3} presents the $p_T$ dependences of hadron-hadron charge azimuthal correlations for the centrality bin of 30-50\% in Pb+Pb collisions at 2.76 TeV, where the experimental data are shown by two bands. We can see that the original AMPT results without initial charge separation (circles) can reproduce the experimental opposite-charge data, but fails to describe the same-charge data. The AMPT results with initial two-flavor charge separation (triangles) also can describe the opposite-charge data, but underestimate the magnitudes of same-charge data. The AMPT results with initial three-flavor charge separation can describe the opposite-charge data, but overestimate the same-charge data. Therefore, neither two-flavor case nor three-flavor case can well reproduce the experimental data.

\begin{figure}
\includegraphics[scale=0.6]{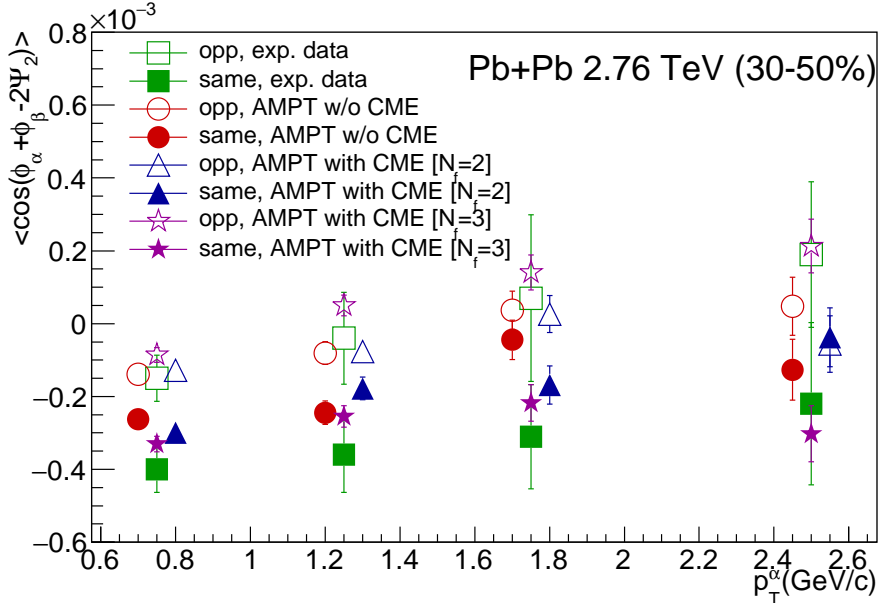}
\caption{(Color online)  Same as Figure~\ref{fig-3} but for kaon-hadron charge azimuthal correlation. 
}
 \label{fig-4}
\end{figure}

We next study the kaon-hadron charge azimuthal correlation, because it may reflect some sensitive information about the flavor dependence of the CME without any influence from $\rho$ meson decays. Figure~\ref{fig-4} shows the $p_T$ dependences of kaon-hadron charge azimuthal correlations for the centrality bin of 30-50\% in Pb+Pb collisions at 2.76 TeV, where squares represent the new ALICE data. It seems that the original AMPT results without initial charge separation (circles) can only describe experimental opposite-charge data, but overestimate the same-charge data. It appears that both two-flavor case and three-flavor case can describe the experimental data if one considers the large experimental errors. It is difficult to judge which of the two CME cases is better for the moment. On the other hand, since the difference between two-flavor and three-flavor results is limited, it will be helpful to have more event statistics to disentangle two scenarios of CME clearly in future experiments.

\section{Summary}
\label{sec:summary}

We introduce a centrality-dependent initial two-flavor and three-flavor charge separation into the AMPT model which is a very good background/baseline model to study the quark flavor dependence of the CME. We observe that both kinds of CME can improve the description to the traditional  CME observable of charge azimuthal correlation $\gamma$. We compare the AMPT results with the new ALICE measurement of the charge azimuthal correlations with identified hadrons (i.e. hadron-hadron correlation vs. kaon-hadron correlation). We find that the charge azimuthal correlation with identified hadrons, especially with identified kaons, has a potential to disentangle two-flavor CME from three-flavor CME. However, within the current experimental uncertainties, it is difficult to conclude which-flavor CME the ALICE data favor. Therefore, more statistics are required to verify this clearly in future experiments, which would help us to further understand more aspects of the QCD phase transitions.

\section*{ACKNOWLEDGMENTS}

We thank Jin-Hui Chen, Heng-Tong Ding and Jacobus Onderwaater for their helpful discussions. We are grateful to Chen Zhong for maintaining the computing resources. This work is supported by the Major State Basic Research Development Program in China under Grant No. 2014CB845404, the National Natural Science Foundation of China under Grants No. 11522547, No. 11375251, and No. 11421505 and the Program for the Excellent Youth at Henan Normal University under Grant No. 154100510007.

\end{document}